\documentclass[a4paper,10pt]{scrartcl}
\pdfoutput=1
\usepackage[utf8]{inputenc}
\usepackage[T1]{fontenc}
\usepackage[english]{babel}
\usepackage[binary-units=true]{siunitx}
\usepackage{amsmath,amssymb,array}
\usepackage{booktabs}
\usepackage{csquotes}
\usepackage[round]{natbib}
\usepackage{graphicx}
\usepackage{hyperref}
\usepackage{hyphenat}
\usepackage{microtype}
\usepackage{verbatim}
\usepackage[inline,shortlabels]{enumitem}

\newcommand{\CRANpkg}[1]{\texttt{\href{https://cran.r-project.org/package=#1}{#1}}}
\newcommand{\pkg}[1]{\texttt{#1}}
\newcommand{\R}{\texttt{R}}
\newcommand{\code}[1]{\texttt{#1}}
\title{checkmate: Fast Argument Checks for Defensive \R{} Programming}
\author{Michel Lang\\\url{lang@statistik.tu-dortmund.de}}
\date{}

\begin{document}

\maketitle

\abstract{%
  Dynamically typed programming languages like \R{} allow programmers to write generic, flexible and concise code and to interact with the language using an interactive Read-eval-print-loop (REPL).
  However, this flexibility has its price: As the \R{} interpreter has no information about the expected variable type, many base functions automatically convert the input instead of raising an exception.
  Unfortunately, this frequently leads to runtime errors deeper down the call stack which obfuscates the original problem and renders debugging challenging.
  Even worse, unwanted conversions can remain undetected and skew or invalidate the results of a statistical analysis.
  As a resort, assertions can be employed to detect unexpected input during runtime and to signal understandable and traceable errors.
  The package \CRANpkg{checkmate} provides a plethora of functions to check the type and related properties of the most frequently used \R{} objects and variable types.
  The package is mostly written in C to avoid any unnecessary performance overhead.
  Thus, the programmer can conveniently write concise, well-tested assertions which outperforms custom \R{} code for many applications.
  Furthermore, checkmate simplifies writing unit tests using the framework~\CRANpkg{testthat}~\citep{wickham_2011} by extending it with plenty of additional expectation functions, and registered C routines are available for package developers to perform assertions on arbitrary SEXPs (internal data structure for \R{} objects implemented as struct in C) in compiled code.
}

\section{Defensive Programming in \R{}}
\label{sec:introduction}
Most dynamic languages utilize a weak type system where the type of variable must not be declared, and \R{} is no exception in this regard.
On the one hand, a weak type system generally reduces the code base and encourages rapid prototyping of functions.
On the other hand, in comparison to strongly typed languages like C/C++, errors in the program flow are much harder to detect.
Without the type information, the \R{} interpreter just relies on the called functions to handle their input in a meaningful way.
Unfortunately, many of \R{}'s base functions are implemented with the REPL in mind.
Thus, instead of raising an exception, many functions silently try to auto-convert the input.
E.g., instead of assuming that the input \code{NULL} does not make sense for the function \code{mean()}, the value \code{NA} of type numeric is returned and additionally a warning message is signaled.
While this behaviour is acceptable for interactive REPL usage where the user can directly react to the warning, it is highly unfavorable in packages or non-interactively executed scripts.
As the generated missing value is passed to other functions deeper down the call stack, it will eventually raise an error.
However, the error will be reported in a different context and associated with different functions and variable names.
The link to origin of the problem is missing and debugging becomes much more challenging.
Furthermore, the investigation of the call stack with tools like \code{traceback()} or \code{browser()} can result in an overwhelming number of steps and functions.
As the auto-conversions cascade nearly unpredictably (as illustrated in Table~\ref{tab:ex_base_funs}), this may lead to undetected errors and thus to misinterpretation of the reported results.

\begin{table}[ht]
  \footnotesize
  \centering
  \begin{tabular}{l|cccc}\toprule
                           & \multicolumn{4}{c}{Return value of} \\
    Input                  & \code{mean(x)}        & \code{median(x)}       & \code{sin(x)}     & \code{min(x)}          \\ \midrule
    \code{numeric(0)}      & \code{NaN}            & \code{NA}              & \code{numeric(0)} & \code{Inf} (w)         \\
    \code{character(0)}    & \code{NA\_real\_} (w) & \code{NA\_character\_} & [exception]       & \code{NA\_character\_} (w) \\
    \code{NA}              & \code{NA\_real\_}     & \code{NA}              & \code{NA\_real\_} & \code{NA\_integer\_}   \\
    \code{NA\_character\_} & \code{NA\_real\_} (w) & \code{NA\_character\_} & [exception]       & \code{NA\_character\_} \\
    \code{NaN}             & \code{NaN}            & \code{NA}              & \code{NaN}        & \code{NaN}             \\
    \code{NULL}            & \code{NA} (w)         & \code{NULL} (w)        & [exception]       & \code{Inf} (w)         \\
    \bottomrule
  \end{tabular}
  \caption{Input and output for some simple mathematical functions from the \code{base} package (\R{}-3.3.1).
    Outputs marked with \enquote{(w)} have issued a warning message.
  }\label{tab:ex_base_funs}
\end{table}

The described problems lead to a concept called \enquote{defensive programming} where the programmer is responsible for manually checking function arguments.
Reacting to unexpected input as soon as possible by signaling errors instantaneously with a helpful error message is the key aspect of this programming paradigm.
A similar concept is called \enquote{design by contract} which demands the definition of formal, precise and verifiable input and in return guarantees a sane program flow if all preconditions hold.
The package \pkg{checkmate} assists the programmer in writing such assertions in a concise way for the most important \R{} objects.

\section{Related work}
\label{sec:related_work}
Many packages contain custom code to perform argument checks.
These either rely on (a) the base function \code{stopifnot()} or (b) hand-written cascades of \code{if-else} blocks containing calls to \code{stop()}.
Option (a) can be considered a quick hack because the raised error messages lack helpful details or instructions for the user.
Option (b) is the natural way of doing argument checks in \R{} but quickly becomes tedious.
For this reason many packages have their own functions included, but there are also some packages on CRAN whose sole purpose are argument checks.

The package~\CRANpkg{assertthat}~\citep{wickham_2013} provides the \enquote{drop-in replacement} \code{assert\_that()} for \R{}'s \code{stopifnot()} while generating more informative help messages.
This is achieved by evaluating the expression passed to the function \code{assert\_that()} in an environment where functions and operators from the base package (e.g.\, \code{as.numeric()} or
\code{`==`}) are overloaded by more verbose counterparts.
E.g., to check a variable to be suitable to pass to the \code{log()} function, one would require a numeric vector with all positive elements and no missing values:
\begin{verbatim}
assert_that(is.numeric(x), length(x) > 0,
  all(!is.na(x)), all(x >= 0))
\end{verbatim}
Furthermore, \pkg{assertthat} offers some additional convenience functions like \code{is.flag()} to check for single logical values or \code{has\_name()} to check for presence of specific names.
These functions also prove useful if used  with \code{see\_if()} instead of \code{assert\_that()} which turns the passed expression into a predicate function returning a logical value.

The package \CRANpkg{assertive}~\citep{cotton_2016} is another popular package for argument checks.
Its functionality is split over 15~packages containing over 400~functions, each specialized for a specific class of assertions:
For instance, \CRANpkg{assertive.numbers} specialises on checks of numbers and \CRANpkg{asserive.sets} offers functions to work with sets.
The functions are grouped into functions starting with \code{is\_} for predicate functions and functions starting with \code{assert\_} to perform \code{stopifnot()}-equivalent operations.
The author provides a \enquote{checklist of checks} as package vignette to assist the user in picking the right functions for common situations like checks for numeric vectors or for working with files.
Picking up the \code{log()} example again, the input check with \pkg{assertive} translates to:
\begin{verbatim}
assert_is_numeric(x)
assert_is_non_empty(x)
assert_all_are_not_na(x)
assert_all_are_greater_than_or_equal_to(x, 0)
\end{verbatim}

Moreover, the package \CRANpkg{assertr}~\citep{fischetti_2016} focuses on assertions for \CRANpkg{magrittr}~\citep{bache_2014} pipelines and data frame operations in \CRANpkg{dplyr}~\citep{wickham_2016}, but is not intended for generic runtime assertions.

\section{The \pkg{checkmate} Package}
\label{sec:checkmate}

\subsection{Design goals}
\label{ssec:design}
The package has been implemented with the following goals in mind:
\begin{description}
  \item[Runtime]
    To minimize any concern about the extra computation time required for assertions, most functions directly jump into compiled code to perform the assertions directly on the SEXPs.
    The functions also have been extensively optimized to first perform inexpensive checks in order to be able to skip the expensive ones.
  \item[Memory] In many domains the user input can be rather large, e.g.\ long vectors and high dimensional matrices are common in text mining and bioinformatics.
    Basic checks, e.g.\ for missingness, are already quite time consuming, but if intermediate objects of the same dimension have to be created, runtimes easily get out of hand.
    For example, \code{any(x <\ 0)} with \code{x} being a large numeric matrix internally first allocates a logical matrix \code{tmp} with the same dimensions as \code{x}.
    The matrix \code{tmp} is then passed in a second step to \code{any()} which aggregates the logical matrix to a single logical value and \code{tmp} is marked to be garbage collected.
    Besides a possible shortage of available memory, which may cause the machine to swap or the R interpreter to terminate, runtime is wasted with unnecessary memory management.
    \pkg{checkmate} solves this problem by looping directly over the elements and thereby avoiding any intermediate objects.
  \item[Code completion]
    The package aims to provide a single function for all frequently used \R{} objects and their respective characteristics and attributes.
    For example, the function \code{assertNumeric()} provides arguments to check for length, missingness and lower/upper bound.
    After typing the function name, the code completion of editors which speak \R{} can suggest additional checks for the respective variable type.
    This context-sensitive assistance often helps writing more concise assertions.
\end{description}

\subsection{Naming scheme}
\label{ssec:naming}
The core functions of the package follow a specific naming scheme:
The first part (prefix) of a function name determines the action to perform w.r.t.\ the outcome of the respective check while the second part of a function name (suffix) determines the base type of the object to check.
The first argument of all functions is always the object~\code{x} to check and further arguments specify additional restrictions on \code{x}.

\subsubsection{Prefixes}
There are currently four families of functions, grouped by their prefix, implemented in \pkg{checkmate}:
\begin{description}
  \item[assert*] Functions prefixed with \enquote{assert} throw an exception if the corresponding check fails and the checked object is returned invisibly on success.
    This family of functions is suitable for many different tasks.
    Besides argument checks of user input, this family of functions can also be used as a drop-in replacement for \code{stopifnot()} in unit tests using the internal test mechanism of \R{} as described in Writing R Extensions \citep{rcoreteam_2016}, subsection 1.1.5.
    Furthermore, as the object to check is returned invisibly, the functions can also be used inside \pkg{magrittr} pipelines.
  \item[test*] Functions prefixed with \enquote{test} are predicate functions which return \code{TRUE} if the respective check is successful and \code{FALSE} otherwise.
    This family of functions is best utilized if different checks must be combined in a non-trivial manner or custom error messages are required.
  \item[expect*] Functions prefixed with \enquote{expect} are intended to be used together with \pkg{testthat}: the check is translated to an expectation which is then forwarded to the active \pkg{testthat} reporter.
    This way, \pkg{checkmate} extends the facilities of \pkg{testthat} with dozens of powerful helper functions to write efficient and comprehensive unit tests.
    Note that \pkg{testthat} is an optional dependency and the \code{expect}-functions only work if \pkg{testthat} is installed.
    Thus, to use \pkg{checkmate} as an \pkg{testthat} extension, \pkg{checkmate} must be listed in \code{Suggests} or \code{Imports} of a package.
  \item[check*] Functions prefixed with \enquote{check} return the error message as a string if the respective check fails, and \code{TRUE} otherwise.
    Functions with this prefix are the workhorses called by the \enquote{asssert}, \enquote{test} and \enquote{expect} families of functions and prove especially useful to implement custom assertions.
    They can also be used to collect error messages in order to generate reports of multiple check violations at once.
\end{description}

The prefix and the suffix can be combined in both \enquote{camelBack} and \enquote{underscore\_case} fashion.
In other words, checkmate offers all functions with the \enquote{assert}, \enquote{test} and \enquote{check} prefix in both programming style flavors: \code{assert\_numeric()} is a synonym for \code{assertNumeric()} the same way \code{testDataFrame()} can be used instead of \code{test\_data\_frame()}.
By supporting the two most predominant coding styles for \R{}, most programmers can stick to their favorite style while implementing runtime assertions in their packages.

\subsubsection{Suffixes}
While the prefix determines the action to perform on a successful or failed check, the second part of each function name defines the base type of the first argument~\code{x}, e.g.\ \code{integer}, \code{character} or \code{matrix}.
Additional function arguments restrict the object to fulfill further properties or attributes.

\paragraph{Atomics and Vectors}
The most important built-in atomics are supported via the suffixes \code{*Logical}, \code{*Numeric}, \code{*Integer}, \code{*Complex}, \code{*Character}, \code{*Factor}, and \code{*List} (strictly speaking, \enquote{numeric} is not an atomic type but a naming convention for objects of type \code{integer} or \code{double}).
Although most operations that work on real values also are applicable to natural numbers, the contrary is often not true.
Therefore numeric values frequently need to be converted to integer, and \code{*Integerish} ensures a conversion without surprises by checking double values to be \enquote{nearby} an integer w.r.t.\ a machine-dependent tolerance.
Furthermore, the object can be checked to be a \code{vector}, an \code{atomic} or an atomic vector (a \code{vector}, but not \code{NULL}).

All functions can optionally test for missing values (any or all missing), length (exact, minimum and maximum length) as well as names being
\begin{enumerate*}[(a)]
  \item not present,
  \item present and not \code{NA}/empty,
  \item present, not \code{NA}/empty and unique, or
  \item present, not \code{NA}/empty, unique and additionally complying to \R{}'s variable naming scheme.
\end{enumerate*}
There are more type-specific checks, e.g.\ bound checks for numerics or regular expression matching for characters.
These are documented in full detail in the manual.

\paragraph{Scalars}
Atomics of length one are called scalars.
Although \R{} does not differentiate between scalars and vectors internally, scalars deserve particular attention in assertions as arguably most function arguments are expected to be scalar.
Although scalars can also be checked with the functions that work on atomic vectors and additionally restricting to length~1 via argument \code{len}, \pkg{checkmate} provides some useful abbreviations: \code{*Flag} for logical scalars, \code{*Int} for an integerish value, \code{*Count} for a non-negative integerish values, \code{*Number} for numeric scalars and \code{*String} for scalar character vectors.
Missing values are prohibited for all scalar values by default as scalars are usually not meant to hold data where missingness occurs naturally (but can be allowed explicitly via argument \code{na.ok}).
Again, additional type-specific checks are available which are described in the manual.

\paragraph{Compound types}
The most important compound types are matrices/arrays (vectors of type logical, numeric or character with attribute \code{dim}) and data frames (lists with attribute \code{row.names} and class \code{data.frame} storing atomic vectors of same length).
The package also includes checks for the popular \code{data.frame} alternatives \CRANpkg{data.table}~\citep{dowle_2014} and \CRANpkg{tibble}~\citep{wickham_2016-1}.
Some checkable characteristics conclude the internal type(s), missingness, dimensions or dimension names.

\paragraph{Miscellaneous}
On top of the already described checks, there are functions to work with sets (\code{*Subset}, \code{*Choice} and \code{*SetEqual}), environments (\code{*Environment}) and objects of class \enquote{Date} (\code{*Date}).
The \code{*Function} family checks \R{} functions and its arguments and \code{*OS} allows to check if R is running on a specific operating system.
The functions \code{*File} and \code{*Directory} test for existence and access rights of files and directories, respectively.
The function \code{*PathForOutput} allows to check whether a directory can be used to store files in it.
Furthermore, \pkg{checkmate} provides functions to check the class or names of arbitrary \R{} objects with \code{*Class} and \code{*Names}.

\paragraph{Custom checks}
Extensions are possible by writing a \code{check*} function which returns \code{TRUE} on success and an informative error message otherwise.
The exported functionals \code{makeAssertionFunction()}, \code{makeTestFunction()} and \code{makeExpectationFunction()} can wrap this custom check function to create the required counterparts in such a way that they seamlessly fit into the package.
The vignette demonstrates this with a check function for square matrices.

\subsection{DSL for argument checks}
\label{ssec:dsl}
Most basic checks can alternatively be performed using an implemented Domain Specific Language (DSL) via the functions \code{qassert()}, \code{qtest()} or \code{qexpect()}.
All three functions have two arguments: The arbitrary object \code{x} to check and a \enquote{rule} which determines the checks to perform provided as a single string.
Each rules consist of up to three parts:
\begin{enumerate}
  \item The first character determines the expected class of \code{x}, e.g.\ \enquote{n} for numeric, \enquote{b} for boolean, \enquote{f} for a factor or \enquote{s} for a string (more can be looked up in the manual).
    By using a lowercase letter, missing values are permitted while an uppercase letter disallows missingness.
  \item The second part is the length definition.
    Supported are \enquote{?} for length~0 or length~1, \enquote{+} for length~$\geq 1$ as well as arbitrary length specifications like \enquote{1}/\enquote{==1} for exact length~1 or \enquote{<10} for length~$<10$.
  \item The third part triggers a range check, if applicable, in interval notation (e.g., \enquote{$[0,1)$} for values $0 \leq x < 1$).
    If the boundary value on an open side of the interval is missing, all values of \code{x} will be checked for being $>-\infty$ or $< \infty$, respectively.
\end{enumerate}
Although this syntax requires some time to familiarize with, it allows to write extensive argument checks with very few keystrokes.
For example, the previous check for the input of \code{log()} translates to the rule \code{"N+[0,]"}.
As the function signature is really simplistic, it is perfectly suited to be used from compiled code written in C/C++ to check arbitrary \code{SEXP}s.
For this reason \pkg{checkmate} provides header files which foreign packages can link against.
Instructions can be found in the package vignette.

\section{Benchmarks}
\label{sec:benchmarks}
This small benchmark study picks up the $\log()$ example once again: testing a vector to be numeric with only positive, non-missing values.

\subsection{Implementations}
\label{ssec:implementations}
Now we compare \pkg{checkmate}'s \code{assertNumeric()} and \code{qassert()} (as briefly described in the previous Section~\nameref{ssec:dsl}) with counterparts written with \R{}'s \code{stopifnot()}, \pkg{assertthat}'s \code{assert\_that()} and a series of \pkg{assertive}'s \code{assert\_*()} functions:
\begin{verbatim}
checkmate <- function(x) {
  assertNumeric(x, any.missing = FALSE, lower = 0)
}

qcheckmate <- function(x) {
  qassert(x, "N[0,]")
}

R <- function(x) {
  stopifnot(is.numeric(x), all(!is.na(x)), all(x >= 0))
}

assertthat <- function(x) {
  assert_that(is.numeric(x), all(!is.na(x)), all(x >= 0))
}

assertive <- function(x) {
  assert_is_numeric(x)
  assert_all_are_not_na(x)
  assert_all_are_greater_than_or_equal_to(x, 0)
}
\end{verbatim}
To allow measurement of failed assertions, the above functions are wrapped into a \code{try()}.
The source code for this benchmark study is provided in the the supplementary.

\subsection{Setup}
The benchmark was performed on an Intel i5-6600 with \SI{16}{\giga\byte} running \R{}-3.3.1 on a 64bit Arch Linux installation.
The package versions are 1.8.2 for \pkg{checkmate}, 0.1 for \pkg{assertthat} and 0.3-4 for \pkg{assertive}.
\R{}, the linked OpenBLAS and all packages have been compiled with the GNU Compiler Collection (GCC) in version 6.2.1 and tuned with \code{march=native} on optimization level \code{-O2}.
To compare runtime differences, \CRANpkg{microbenchmark}~\citep{mersmann_2015} is setup to do 100~replications.
The wrappers have also been compared to their byte-compiled version (using \code{compiler::cmpfun}) with no notable difference in performance, thus the later presented results are extracted from the uncompiled versions of these wrappers.

\subsection{Results}
\begin{figure}[ht]
  \centering
  \includegraphics[width=0.99\textwidth]{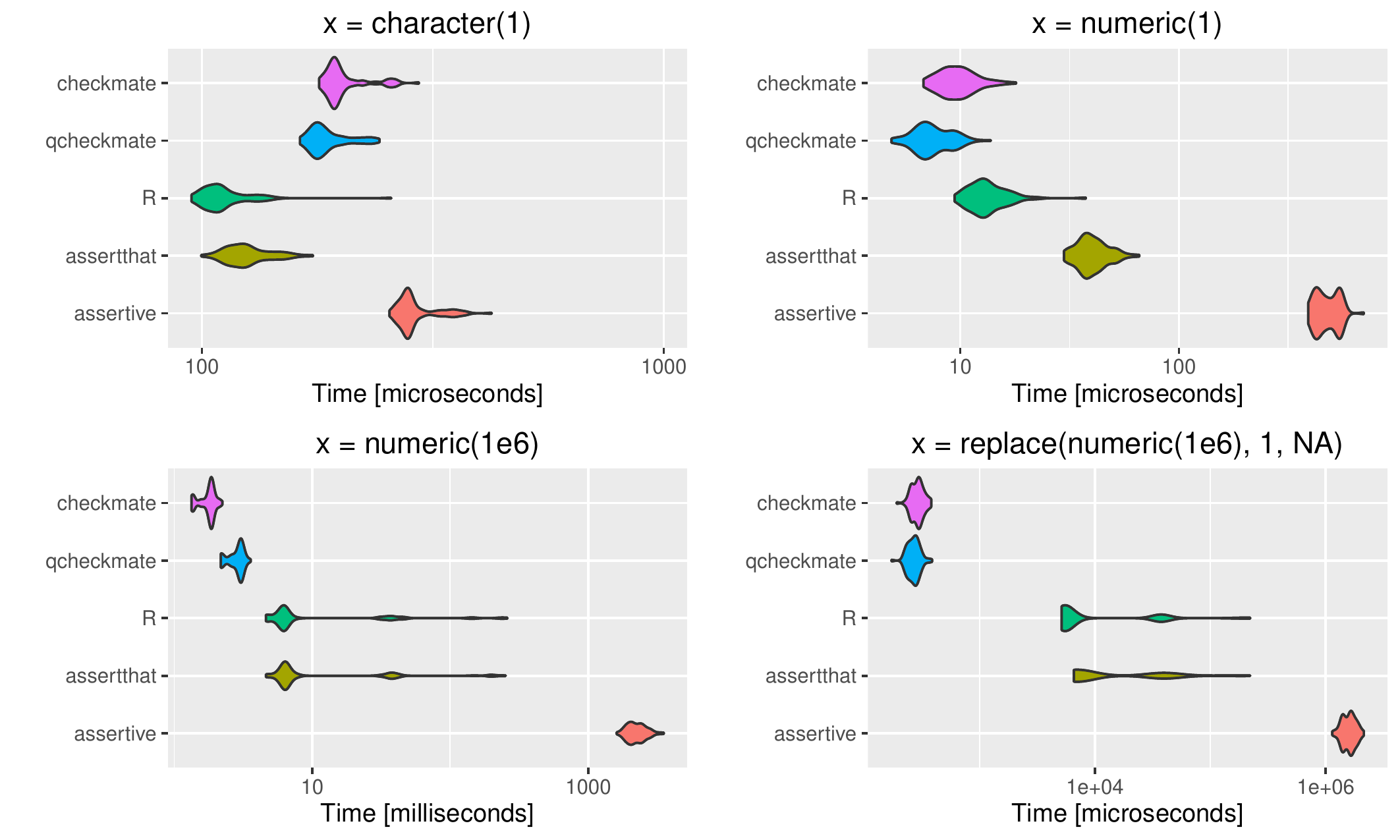}
  \caption{Violin plots of the runtimes on $\log_{10}$-scale of the assertion \enquote{$x$ must be a numeric vector with all elements positive and no missing values} on different input \code{x}.}
  \label{fig:benchmark}
\end{figure}

The benchmark is performed on four different inputs and the resulting timings are presented in Figure~\ref{fig:benchmark}.
Note that the runtimes on the $x$-axis are on $\log_{10}$-scale and use different units of measurement.

\begin{description}
  \item[top left] Input \code{x} is a scalar character value, i.e.\ of wrong type.
    This benchmark serves as a measurement of overhead: the first performed and cheapest assertion on the type of \code{x} directly fails.
    In fact, all assertion frameworks only require microseconds to terminate.
    \R{} directly jumps into compiled code via a \code{Primitive} and therefore has the least overhead.
    \pkg{checkmate} on the other hand has to jump into the compiled code via the \code{.Call} interface which is comparably slower.
    The implementation in \pkg{assertthat} is faster than \pkg{checkmate} (as it also primarily calls primitives) but slightly slower than \code{stopifnot()}. The implementation in \pkg{assertive} is the slowest.
    However, in case of assertions (in comparison to tests returning logicals), the runtimes for a successful check are arguably more important than for a failed check because the latter raises an exception which usually is a rare event in the program flow and thus is not time-critical.
    Therefore, the next benchmark might be more relevant for many applications.
  \item[top right] Input \code{x} is a scalar numeric value.
    The implementations now additionally check for missingness and negative values and do not raise an exception.
    \code{qassert()} is the fastest implementation, followed by \code{assertNumeric()}.
    Although \code{qassert()} and \code{assertNumeric()} basically call the same code internally, \code{qassert()} has less overhead due to its minimalistic interface.
    \R{}'s \code{stopifnot()} is a tad slower comparing the median runtimes but still faster than \pkg{assertthat} (5x slowdown in comparison to \code{qassert()}).
    \pkg{assertive} is >60x slower than \code{qassert()}.
  \item[bottom left] Input \code{x} is now a long vector with $10^6$ numeric elements.
    \pkg{checkmate} has the fastest versions with a speedup of approximately 3.5x compared to \R{}'s \code{stopifnot()} and \code{assert\_that()}.
    In comparison to its alternatives, \pkg{checkmate} avoids intermediate objects as described in \nameref{ssec:design}: Instead of allocating a \code{logical(1e6)} vector first to aggregate it in a second step, \pkg{checkmate} directly operates on the numeric input.
    That is also the reason why \code{stopifnot()} and \code{assertthat()} have high variance in their runtimes: The garbage collector occasionally gets triggered to free memory which requires a substantial amount of time.

    \pkg{assertive} is orders of magnitude slower for this input (>1200x) because it follows a completely different philosophy: Instead of focusing on speed, \pkg{assertive} gathers detailed information while performing the assertion.
    This yields report-like error messages (e.g., the index and reason why an assertion failed, for each element of the vector) but is comparably slow.
  \item[bottom right] Input \code{x} is again a large vector, but the first element is a missing value.
    Here, all implementations first successfully check the type of \code{x} and then throw an error about the missing value.
    Again, \pkg{checkmate} avoids allocating intermediate objects which in this case yields an even bigger speedup:
    While the other packages first check $10^6$ elements for missingness to create a \code{logical(1e6)} vector which is then passed to \code{any()}, \pkg{checkmate} directly stops after analyzing the first element of \code{x}.
    This obvious optimization yields a speedup of 20x in comparison to \R{} and \pkg{assertthat} and a 6000x speedup in comparison to \pkg{assertive}.
\end{description}

Summed up, \pkg{checkmate} is the fastest option to perform expensive checks and only causes a small decrease in performance for trivial, inexpensive checks which fail quickly (top left).
Although the runtime differences seem insignificant for small input (top right), the saved microseconds can easily sum up to seconds or hours if the respective assertion is located in a hot spot of the program and therefore is called millions of times.
For large input, the runtime differences are often notable without benchmarks, and even become much more important as data grows bigger.

\section{Conclusion}
\label{sec:conclusion}
Runtime assertions are a necessity in \R{} to ensure a sane program flow, but \R{} itself offers very limited capabilities to perform these kind of checks.
\pkg{checkmate} allows programmers and package developers to write assertions in a concise way without unnecessarily sacrificing runtime performance nor increasing the memory footprint.
Compared to the presented alternatives, assertions with \pkg{checkmate} are faster, tailored for bigger data and (with the help of code completion) more convenient to write.
They generate helpful error messages, are extensively tested for correctness and suitable for large and extensive software projects (\CRANpkg{mlr}~\citep{bischl_2016} and \CRANpkg{BatchJobs}~\citep{bischl_2015} already make heavy use of \pkg{checkmate}).
Furthermore, \pkg{checkmate} offers capabilities to do assertions on SEXPs in compiled code via a domain specific language and extends the popular unit testing framework \pkg{testthat} with many helpful expectation functions.

\section{Acknowledgments}
Part of the work on this paper has been supported by Deutsche Forschungsgemeinschaft (DFG) within the Collaborative Research Center SFB~876 \enquote{Providing Information by Resource-Constrained Analysis}, project~A3 (\href{http://sfb876.tu-dortmund.de}{http://sfb876.tu-dortmund.de}).

\bibliography{RJreferences}

\end{document}